\documentclass[fleqn,twoside]{article}
\usepackage{espcrc2}

\usepackage{graphicx}

\newcommand{\AmS}{{\protect\the\textfont2
  A\kern-.1667em\lower.5ex\hbox{M}\kern-.125emS}}

\title{ AdS space compactification and holographic mapping in   
the AdS/CFT correspondence }

\author{Henrique Boschi-Filho\address[UFRJ]{Instituto de F\'\i sica, 
Universidade Federal do Rio de Janeiro, 
Caixa Postal 68528, 21941-972,  Rio de Janeiro, RJ, Brazil}\thanks{boschi@if.ufrj.br} 
and Nelson R. F. Braga\addressmark[UFRJ]\thanks{braga@if.ufrj.br}}

\begin{document}
  
\begin{abstract}
Physical consistency of quantum fields in anti-de Sitter space time requires
that the space must be compactified by the  inclusion
of a boundary where appropriate conditions are imposed. 
An interpretation for the presence of this boundary 
is found taking AdS as a limiting case of the space generated by a large number of 
coincident branes. 
The compactification of AdS leads to a discretization of the spectrum of bulk fields. 
As a consequence, we find a one to one mapping between the quantum states
of scalar fields in AdS bulk and boundary. 
Using this mapping as an approximation for the dual relation between
string dilaton field and scalar QCD glueballs the high energy QCD scaling 
is reproduced. We also use this map to estimate the ratio of 
scalar glueball masses.
\end{abstract}

\maketitle
 
\section{Introduction}

According to the Maldacena conjecture\cite{Malda} the large $N$ limit of $SU(N)$
superconformal field theories in $n$ dimensions can be described by supergravity in
$n+1$ dimensional anti de Sitter (AdS) spacetime times a compact manifold. 
This is known as the AdS/CFT correspondence.  
In this correspondence (see also refs. \cite{GKP,Wi,MV,FMMR} and for reviews see
\cite{Malda2,Pe,Kle}) 
the AdS space shows up  as a near horizon geometry of a set of
coincident D3-branes. A Dirichlet $p$-brane or D$p$-brane is a $p+1$ dimensional 
hyperplane where strings are allowed to end\cite{HS,Po}).
The AdS/CFT correspondence can be understood as a realisation of the Holographic 
principle: "the degrees of freedom of  a quantum theory with gravity can be 
mapped on the corresponding 
boundary"\cite{HOL1,HOL2,HOL3,HOL4,HOL5}. 

Precise prescriptions for the realization of the AdS/CFT correspondence 
were presented in \cite{GKP,Wi} by considering  Poincare patches of AdS space.
Originally AdS space was formulated using global coordinates. However
Poincare coordinate system allows
a simple definition for the flat boundary where the conformal field theory 
is defined as we will discuss in the next section.  In section {\bf 3} we 
discuss the quantization of scalar bulk fields in AdS space which allow us 
to write a one to one mapping between bulk and boundary scalar fields. 
In section {\bf 4} we apply this holographic mapping to describe high energy QCD
scaling of scalar glueballs. We also comment on the possibility of applying
this holographic mapping to obtain the ratio of scalar glueball masses.

\section{AdS space and compactification}
AdS space can be described by global coordinates with finite ranges\cite{HE}.
In this context it was shown\cite{QAdS1,QAdS2} that a consistent quantization 
can only be obtained if  one  supplements AdS space 
with a boundary  in order to impose vanishing
flux of particles and  information there.
Otherwise massless particles would be able to go to (or come from)  
spatial infinity in finite times and it would thus be impossible 
to define a Cauchy surface. 

In the AdS/CFT correspondence one considers the AdS space represented by 
Poincar\'e coordinates $\,z \,,\,\vec x\,,\,t\,$  with metric
\begin{equation}
\label{metric}
ds^2=\frac {R^2 }{( z )^2}\Big( dz^2 \,+(d\vec x)^2\,
- dt^2 \,\Big)\,.
 \end{equation}

\noindent The boundary in this case is given by the surface  
$z\,=\,0\,$ plus a point defined by the limit 
$\,z\,\rightarrow\,\infty\,$.
Thus a consistent quantization requires the inclusion of
this surface and the extra point at infinity where the flux should be 
required to vanish.
This is not possible using just one set of Poincar\'e coordinates.
The point at infinity can not be accommodated in the original Poincare 
chart so that  we have to introduce a second coordinate system
$\,z^\prime \,,\,\vec x\,,\,t\,$\cite{BB1}.
The point $z \rightarrow \infty$  can be represented 
in the second chart at some positive constant value $z^\prime\,=\,\delta $.
The coordinates $z$ and $z^\prime $ of the two charts are related in this case by
\begin{equation}
{1\over z^\prime} = {1\over \delta} - {1\over z}\,, 
\end{equation}
 
\noindent with range $\delta \le z^\prime < \infty \,$.
The metric of the second coordinate system involves a Poincare like factor
$( R/z^{\prime})^2\,$ and can be written as\cite{BB2}
\begin{equation}
\label{metric2}
ds^2 = {R^2 \over z^{\prime\,2}} \Big\lbrack
{ \delta^2 dz^{\prime\,2}\over (z^\prime - \delta )^2 } +
{ (z^\prime - \delta )^2 \over \delta^2 } \Big( (d\vec x)^2
- dt^2 \Big)\Big\rbrack\,.
\end{equation}

Now the whole compact AdS space is described by the coordinate charts corresponding to 
eqs. (\ref{metric}) and (\ref{metric2}). 
Further, with this second chart we find a horizon 
(infinite singularity in the spatial part of  $ds^2$) 
at $z^\prime = \delta$. This was not apparent in the original Poincare chart.
Note that there is another horizon at $ z = 0$. We will comment on this latter.

Let us now discuss the ten dimensional geometry generated by $N$ coincident D3-branes
and its relation to the compactified AdS space. 
The D3-brane metric can be written as\cite{HS,GKP} 
\begin{eqnarray}
\label{branemetric}
ds^2 &=& \Big( 1 + {R^4\over r^4} \Big)^{-1/2} ( -dt^2 + d{\vec x}^2 ) \nonumber\\  
&+& \Big( 1 + {R^4\over r^4} \Big)^{1/2} (dr^2 + r^2 d\Omega^2_5 )\,,
\end{eqnarray}

\noindent where we are using the same symbol $R$ for 
a constant that now satisfies $R^4 \,=\, N/ 2\pi^2 T_3$ where
$T_3$ is the tension of a single D3-brane. 
The metric (\ref{branemetric}) has  a horizon at $r = 0$ with zero perpendicular area 
(apart from the $S_5$ term).
It is interesting to look at the space corresponding to (\ref{branemetric}) in two 
limiting cases where it assumes simpler asymptotic forms: large and small $r$ 
compared to $R$.
Considering first the region $r >>R$ (far from the horizon) the space is 
asymptotically a ten dimensional Minkowski space:
\begin{equation}
\label{flat}
(ds^2)_{far} \,=\, -dt^2 + d{\vec x}^2  +  
dr^2 + r^2 d\Omega^2_5 \,.
\end{equation}
\noindent Now looking at the near horizon region $r << R $ we can approximate 
the metric (\ref{branemetric}) as:
\begin{equation}
ds^2 = {r^2 \over R^2} ( -dt^2 + d{\vec x}^2 ) +  
{R^2\over r^2}dr^2 + R^2 d\Omega^2_5 \,.
\end{equation}
\noindent Changing the axial coordinate according to: $ z = R^2/r$,
as in ref.\cite{Malda,GKP}, 
the metric that describes the brane system for
$ r/ R << 1$ takes the form
\begin{equation}
\label{metric3}
ds^2=\frac {R^2 }{ z^2}\Big( dz^2 \,+(d\vec x)^2\,
- dt^2 \,\Big)\,+\,R^2 d\Omega^2_5\,,
\end{equation}
\noindent corresponding to AdS$_5\times$ S$^5\,$. 
This corresponds to the Poincar\'e chart (\ref{metric}) apart from the $S_5$ factor.
Note however that the horizon  $r=0$ which corresponds to the limit  
$z \rightarrow \infty $
is not included in this chart as a consequence of the lack of a relation between $z$ 
and $r$ at $r=0$. It is interesting to note that from the point of view of a pure AdS
space, one has to include this point as a 
requirement for  a consistent quantization.  Considering the  brane space this point 
is already present,  corresponding to the brane location. 
The inclusion of this point in the AdS space  is possible by introducing one 
more coordinate chart as discussed before.  So, indeed the horizon found in 
the second chart at
$z^\prime = \delta$ corresponds to the brane horizon.

Let us now examine the large $r$ region of the D3-branes space. A  massless particle
moving in the $r \rightarrow \infty $ direction will arrive at an asymptotically
Minkowski space as in eq.(\ref{flat}).
Then it would spend an infinite time to reach spatial infinity.
So, the Cauchy problem is well posed for the D3-branes space
which is geodesically complete. 

Further it is interesting to consider the limit $R \rightarrow \infty $
as suggested by the Maldacena conjecture\cite{Malda}. 
The larger we take $R $ the larger is the range 
of $r$ for which the AdS approximation (\ref{metric3}) for the brane metric 
(\ref{branemetric})  holds. 
So one could naively disregard the asymptotic flat space region in this limit. 
Then one would find an AdS space
without the boundary, where particles could enter or leave the space in finite times.
This would lead to the absence of a well defined Cauchy problem.
 
If one chooses to disregard the flat space region, boundary conditions
should be imposed at $r\rightarrow \infty $ in order to recover
physical consistency.
That means, in the limit $R \rightarrow \infty $ we should not
represent the branes space by just a pure AdS space but rather by a 
compactified AdS including the hypersurface at $z = 0$ besides the point
$z$ at infinity. 
 
It is interesting to note that if we consider the whole space to be of the AdS form
(eq.  (\ref{metric3})) there is a horizon with infinite area at  $z = 0$. 
This is not present in the D3-branes model and it is a consequence of closing the 
AdS space as required for physical consistency once the asymptotic flat space region
has been removed.  
 
\section{Quantum fields in AdS space and holographic mapping}
Now we are going to consider scalar fields in a compact $n+1\,$ dimensional AdS space
in Poincare coordinates.
As discussed above we must introduce two coordinate charts and impose matching conditions 
on the fields in the region of their intersection. We take the first chart to be defined
as  $ 0 \le z \le z_{max}\,$. We can take $z_{max}$ arbitrarily large and map 
as much of AdS space as we want into this chart but the fact that 
$z_{max}$ is finite implies the 
discretization of the spectrum associated with the coordinate $z$.
In the region $0\,\le\,z\,\le z_{max}\,$ we can introduce the 
quantum fields\cite{BB1}
\begin{eqnarray}
\Phi(z,\vec x,t)&=&\sum_{p=1}^\infty 
\int {d^{n-1}k \over (2\pi)^{n-1}}\nonumber\\
&\times& {z^{n/2} \,J_\nu (u_p z ) \over z_{max} w_p(\vec k ) 
J_{\nu\,+\,1} (u_p z_{max} ) }\nonumber\\
&\times& \lbrace { {\bf a}_p(\vec k )\ }
 e^{-iw_p(\vec k ) t +i\vec k \cdot \vec x}\,
\,+\,\,c.c.\rbrace
\label{QF}
\end{eqnarray}

\noindent where $w_p(\vec k ) \,=\,\sqrt{ u_p^2\,+\,{\vec k}^2}$
 and $u_p$ are such that $J_\nu(u_p z_{max})=0$.
The operators ${\bf a}\,,\,\,{\bf a}^{\dagger}\,$ satisfy the commutation 
relations
$$
\Big[ {\bf a}_p(\vec k ),{\bf a}^\dagger_{p^\prime}({\vec k}^\prime  )
\Big] = 2 (2\pi)^{n-1} w_p(\vec k )   
\delta_{p  p^\prime}\delta^{n-1} (\vec k -
{\vec k}^\prime )  
$$
\begin{equation}
\Big[ {\bf a}_p(\vec k ),{\bf a}_{p^\prime}({\vec k}^\prime  )
\Big] = \Big[ {\bf a}^\dagger_p(\vec k ),
{\bf a}^\dagger_{p^\prime}({\vec k}^\prime  ) \Big]=0\,.
\end{equation}

A consequence of the discretization of the field spectrum is that 
we can find a mapping between bulk and boundary states.
Let us define a scalar field on the $n$ dimensional AdS boundary $ z = 0 $ 

\begin{eqnarray}
\Theta ( \vec x ,t)&=& {1\over (2\pi)^{n-1}}
\int^{\infty}_{-\infty} { d^{n-1} K  \over 2 w(\vec K ) }\nonumber\\
&\times& \lbrace { {\bf b}( \vec K )\ }
 e^{-iw(\vec K ) t +i \vec K \cdot  \vec x }\,
\,+\,\,c.c.\rbrace\,.
\end{eqnarray}

\noindent where $ w(\vec K ) = \sqrt{ {\vec K}^2 + \mu^2}$\, 
and the creation-annihilation operators satisfy the canonical algebra
\begin{equation}
\label{canonical2}
\Big[ {\bf b}( \vec K ),{\bf b}^\dagger ({ \vec K }^\prime  )
\Big]\,=\, 2 (2\pi)^{n-1} w( \vec K ) \delta ( \vec K -
{ \vec K}^\prime )\,.
\end{equation}

Following \cite{BB3}, we introduce a sequence of energy scales 
${\cal E}_1 , \,{\cal E}_2,\,...$
and split the spectrum of the operator $\Theta$.
For $0 \le K \le {\cal E}_1$ we introduce the relations between bulk and boundary
operators
\begin{eqnarray}
K^{\,{n-2}\over 2} \,{\bf b}( K , \phi , \theta_\ell ) 
&=& k^{\,{n-2}\over 2}\,{\bf a}_1 (  k, \phi , \theta_\ell  ) \label{11}\\
K^{\,{n-2}\over 2}\,{\bf b}^\dagger ( K , \phi , \theta_\ell ) 
&=& k^{\,{n-2}\over 2}\,{\bf a}^\dagger_1 (  k, \phi , \theta_\ell    )\,,\label{12}
\end{eqnarray}

\noindent where the moduli of the  momenta are mapped onto each other as
$ k = g_1 (K,\mu)\,$. Requiring  that the canonical commutation relations 
for $\bf a $ and $\bf b$ are consistent with the above mapping we find  
a relation between momenta
\begin{eqnarray}
g_1 (K,\mu) &=& {1\over 2} {u_1^2 C_1(\mu) 
\over ( K + \sqrt{K^2 + \mu^2})}\nonumber\\
&-& { K + \sqrt{K^2 + \mu^2}\over 2C_1(\mu)}.
\label{completo}
\end{eqnarray}
For the other intervals we find similar mappings\cite{BB3}.
This way we find a  one to one mapping between bulk and boundary Fock spaces.

It is interesting to note that if we choose a convenient value of
$C_1(\mu) $ and consider  $\mu \ll K \ll {\cal E}_1$ we can
approximate relation (\ref{completo}) as
\begin{equation}
\label{Kk}
k \,\,\approx\,\, { {\cal E}_1\, \mu  \over 2 \, K}\,.
\end{equation}

\section{High energy QCD scaling and glueball masses}
Using the mapping described above with $n = 4$ we can obtain the high energy scaling of 
QCD glueballs at fixed angles. This can be done by relating the scattering amplitudes
of scalar QCD glueballs with those of dilaton fields in AdS bulk.
The assymptotic scalar glueballs are approximated by free scalar fields on 
the AdS boundary. 

We start with the  $S$ matrix for fixed angle scattering in the bulk (dilaton states)
with 2 particles  in the initial state and $m$ particles in the final state  
\begin{eqnarray} 
S_{Bulk} \,\,\,\,\,\,\,\,\,\,\,\,\,\,\,\,\,\,\,\,\,\,\,\,\,\,\,\,\,\,\,\,\,
\,\,\,\,\,\,\,\,\,\,\,\,\,
\,\,\,\,\,\,\,\,\,\,\,\,\,\,\,\,\,\, \,\,\,\,\,\,\,\,\,\,\,\,\,\,\,\,\,\, 
\,\,\,\,\,\,\,\,\,\,\,\,\,\,\,\,\,\, 
\nonumber\\
= \langle  {\vec k}_3,u_1; ...;{\vec k}_{m+2},u_1\,
;out \vert {\vec k}_1,u_1 
;{\vec k}_2 ,u_1;in\rangle
\nonumber\\
=\langle  0\,\vert {\bf a}_{out} ( {\vec k}_3 )... {\bf a}_{out}
 ({\vec k}_{m+2})   {\bf a}^+_{in }
( {\vec k}_1) {\bf a}^+_{in} ({\vec k}_2) \vert  0 \rangle\,. 
\end{eqnarray}

\noindent  Assuming the energy regime to be 
  $ \mu \,\,\ll K \,\ll \,1/ R\,\ll\,k\,\ll\,\, 1/ \sqrt{\alpha^\prime}\,={\cal E}_1\,,
$ where the supergravity approximation for string theory
holds and then using the mapping (\ref{11},\ref{12}) between 
creation-annihilation operators and the approximate relation (\ref{Kk}) we obtain
\begin{eqnarray}
 S_{Bulk}\times\Big({ k \over K}\Big)^{m + 2 } \,\,\,\,\,\,\,
\,\,\,\,\,\,\,\,\,\,\, \,\,\,\,\,\,\,\,\,\,\,\,\,\,\,\,\,\, 
\,\,\,\,\,\,\,\,\,\,\,\,\,\,\,\,\,\, 
\,\,\,\,\,\,\,\, \,\,\,\,\,\,\,\,\,\,\,\,
\nonumber\\
\sim 
 \langle 0 \vert {\bf b}_{out} ( {\vec K}_3 )... 
{\bf b}_{out} ({\vec K}_{m+2})   
{\bf b}^+_{in }( {\vec K}_1) {\bf b}^+_{in} ({\vec K}_2) \vert 0 \rangle  \nonumber
\end{eqnarray}
\begin{equation}
\sim  
\langle  {\vec K}_3 ... {\vec K}_{m+2},out \,\vert \,{\vec K}_1 ,
{\vec K}_2 , in \rangle K^{(m+2)(d-1) }\,.
\end{equation}

\noindent where the field  $\Theta$ on the boundary has
scaling dimension $d$ and then their $in$ and $out$ states are 
$ \vert \vec K \,\rangle \,\cong\, K^{ 1 - d } {\bf b}^+ (\vec K ) \vert 0 \rangle \,$.
This implies a relation between bulk and boundary S matrix elements
\begin{equation}
S_{Bulk} \, \sim \,  
 S_{Bound.} \,\,\Big( {\sqrt{\alpha^\prime} \over \mu }\Big)^{m+2} \,\, K^{(m+2)(1+d)}\,.
\end{equation}
\noindent Finally the corresponding scattering amplitudes ${\cal M}$ are related by
\begin{eqnarray}
\label{Mb}
{\cal M}_{Bound.} \sim {\cal M}_{Bulk} \,\, S_{Bound.} \,( S_{Bulk})^{-1} 
\Big( { K\over k} \Big)^4 \nonumber\\
\sim  {\cal M}_{Bulk} \,\, K^{8 -  (m+2)(d + 1)  } 
\Big( {\sqrt{\alpha^\prime} \over \mu }\Big)^{2 - m}.
\end{eqnarray}

\noindent In the particular energy regime considered we find
$$
{\cal M}_{Boundary} \,\sim \,K^{4 - \Delta } \,\,\sim \,s^{2  - \Delta/2 }\, \,,
$$
\noindent where $\Delta = ( m + 2) d $ is the total scaling dimension
of the scattering particles associated with glueballs on the four 
dimensional boundary and  $ K \sim \sqrt{ s} $.
This result corresponds to the expected QCD scaling behavior\cite{QCD1,BRO}
and was first reproduced in the AdS/CFT context by  Polchinski and Strassler\cite{PS}.

Extending the idea of this mapping to the case where there is a series of 
scalars $\Theta_i\,$ on the boundary we found a simple estimate
for the ratio of scalar glueball 
masses\cite{BB5} that in four dimensions reads:
$$
\label{QCD4}
{ \mu_i\over \mu_1 }\,=\,{\chi_{2\,,\,i}\over \chi_{2\,,\,1}}\,\,,
$$
where $\mu_1$ is the mass of the state $0^{++}\,$ , $\mu_i$ are the masses
of its excitations and $\chi_{2\,,\,i}$ are the zeros of the Bessel function
$J_2$. The results of this estimate are close to the ones found in \cite{MASSG} 
starting with an AdS-Schwarzschild black hole metric, following Witten's 
suggestion\cite{Wi2}. These results are also in good agreement with lattice 
QCD calculations\cite{LAT1,LAT2}.

More recently we have discussed the fact that this approximation for the ratio
of the masses of scalar glueballs does not depend on the
details of the bulk/boundary mapping discussed above. 
Alternatively it can be found taking an AdS slice as an approximation
for the dual space to a confining gauge theory and identifying the dilaton modes 
with the glueball masses\cite{BB6}. We hope that this discussion can be useful
to describe other processes and states of QCD even in an approximated way.

\section*{Acknowledgments} 
The authors are partially supported by CNPq, FINEP, FAPERJ and CAPES
- Brazilian research agencies.



\end{document}